\documentclass[twocolumn,showpacs,aps,prl,superscriptaddress]{revtex4}

\usepackage{graphicx}
\usepackage{dcolumn}
\usepackage{amsmath}
\usepackage{epsfig}
\usepackage[retainorgcmds]{IEEEtrantools}

\input babarsym

\newcommand{\BABARPubYear}    {12}
\newcommand{\BABARPubNumber}  {013}

\newcommand{\SLACPubNumber} {15071}
\newcommand{\LANLNumber} {1206.2543}

\def\Btonunub{\ensuremath{B^{0}\to} {\rm invisible}\xspace}
\def\Btonunubg{\ensuremath{B^{0}\to} {\rm invisible}\ensuremath{ + \gamma}\xspace}
\def\Bptonunub{\ensuremath{B^{+}\to} {\rm invisible}\xspace}
\def\Bptonunubg{\ensuremath{B^{+}\to} {\rm invisible}\ensuremath{ + \gamma}\xspace}

\def\figurebox#1#2#3{%
    \def\arg{#3}%
    \ifx\arg\empty
    {\hfill\vbox{\hsize#2\hrule\hbox to #2{\vrule\hfill\vbox to #1{\hsize#2\vfill}\vrule}\hrule}\hfill}%
    \else
    {\hfill\epsfbox{#3}\hfill}%
    \fi}

\begin{document}

\preprint{\babar-PUB-\BABARPubYear/\BABARPubNumber} 
\preprint{SLAC-PUB-\SLACPubNumber} 

\begin{flushleft}
\babar-PUB-\BABARPubYear/\BABARPubNumber\\
SLAC-PUB-\SLACPubNumber\\
hep-ex/\LANLNumber\\[10mm]
\end{flushleft}

\title{
{\Large \bf \boldmath
Improved Limits on \Bz Decays to Invisible ($+\gamma$) Final States
}}

%
\author{J.~P.~Lees}
\author{V.~Poireau}
\author{V.~Tisserand}
\affiliation{Laboratoire d'Annecy-le-Vieux de Physique des Particules (LAPP), Universit\'e de Savoie, CNRS/IN2P3,  F-74941 Annecy-Le-Vieux, France}
\author{J.~Garra~Tico}
\author{E.~Grauges}
\affiliation{Universitat de Barcelona, Facultat de Fisica, Departament ECM, E-08028 Barcelona, Spain }
\author{A.~Palano$^{ab}$ }
\affiliation{INFN Sezione di Bari$^{a}$; Dipartimento di Fisica, Universit\`a di Bari$^{b}$, I-70126 Bari, Italy }
\author{G.~Eigen}
\author{B.~Stugu}
\affiliation{University of Bergen, Institute of Physics, N-5007 Bergen, Norway }
\author{D.~N.~Brown}
\author{L.~T.~Kerth}
\author{Yu.~G.~Kolomensky}
\author{G.~Lynch}
\affiliation{Lawrence Berkeley National Laboratory and University of California, Berkeley, California 94720, USA }
% ...
\collaboration{The \babar\ Collaboration}
\noaffiliation

\begin{abstract}
We establish improved upper limits on branching fractions for \Bz decays to final states where the decay products are purely invisible 
(\textit{i.e.}, no observable final state particles) and for final states where the only visible product is a photon.
Within the Standard Model, these decays have branching fractions
that are below the current experimental sensitivity, but various models of physics 
beyond the Standard Model predict significant contributions for these channels.
Using 471 million \BB pairs collected at the \Y4S 
resonance by the \babar\ experiment at the \pep2
$e^{+}e^{-}$ storage ring at the SLAC National Accelerator Laboratory, we 
establish upper limits at the 90\% confidence level of $2.4 \times 10^{-5}$ for the branching 
fraction of \Btonunub and $1.7 \times 10^{-5}$ for the branching fraction of \Btonunubg. 
\end{abstract}

\maketitle
This paper presents updated limits on ``disappearance decays'' of \Bz mesons~\cite{conjugates}, where the
\Bz decay contains no observable final state particles, or such ``invisible'' decay products plus a single photon.
We define invisible decay products here to be electrically neutral particles 
that do not generate a signal in the electromagnetic calorimeter.
These results represent an improvement over the previous limits on these decays, which were based on 19\% of the 
present data sample~\cite{prevprl}.

The rate for invisible $B$ decays is negligibly small within the Standard Model (SM) of particle physics, but 
can be larger in several models of new physics.  
The SM decay $\Bz \to 
\nu\bar{\nu}$, which would give such an invisible experimental signature, is strongly helicity-suppressed
by a factor of order $(m_{\nu}/m_{\Bz})^2$~\cite{BB}
and the resulting branching fraction is necessarily 
well below the range of present experimental observability.  
The SM expectation for the $\Bz \to \nu\bar{\nu}\gamma$ branching fraction is 
predicted to be of order $10^{-9}$, with very little uncertainty from hadronic interactions~\cite{nunugam}. 
An experimental observation of an invisible $(+\;\gamma)$ decay of a \Bz
with current experimental sensitivity
would thus be a clear sign of physics beyond the SM.

A phenomenological model motivated by 
the observation of an anomalous number of dimuon events 
by the NuTeV experiment~\cite{NuTeV}
allows for an invisible \Bz decay to a $\bar{\nu}\chi^0_1$ final state, where $\chi^0_1$ is a neutralino, with a branching fraction in the $10^{-7}$ 
to $10^{-6}$ range~\cite{DDR}.  Also, models with large extra dimensions, which would provide a possible solution to the 
hierarchy problem, can have the effect of producing significant, although small, rates for 
invisible \Bz decays~\cite{ADW,AW,DLP}.

The data used in this analysis were collected with the \babar\ detector at the \pep2 $e^{+}e^{-}$ collider at SLAC.  
The data sample corresponds to a luminosity of 
424~\invfb accumulated at the \Y4S resonance and contains $(471 \pm 3)\times 10^{6}$ \BB pair events.
For background studies we also used 45~\invfb collected at a center-of-mass (CM) energy about 40~\mev below \BB 
threshold (off-peak). 

A detailed description of the \babar\ detector is presented in Ref.~\cite{BABARNIM}. Charged particle
momenta are measured in a tracking system consisting of a five-layer double-sided silicon vertex tracker (SVT)
and a 40-layer hexagonal-cell wire drift chamber (DCH).  The
SVT and DCH operate within a 1.5 T solenoidal field, and have a combined solid angle coverage in the CM frame of 90.5\%.
Photons and long-lived neutral hadrons are detected and their energies are measured in a CsI(Tl) 
electromagnetic calorimeter (EMC), which
has a solid angle coverage in the CM frame of 90.9\%.  Muons 
are identified in the instrumented flux return (IFR).
A detector of internally reflected Cherenkov light (DIRC)
is used for identification of charged kaons and pions. A {\tt GEANT4}~\cite{GEANT4} based Monte Carlo (MC) simulation 
of the \babar\ detector response is
used to optimize the signal selection criteria and evaluate the signal detection efficiency.

The detection of invisible $B$ decays uses the fact that 
$B$ mesons are created in pairs, due to flavor conservation in $e^{+}e^{-}$ interactions.
If one $B$ 
is reconstructed in an event, one can thus infer that another $B$
has been produced.
We reconstruct events in which a \Bz decays to $D^{(*)-}\ell^{+}\nu$ (referred to as the ``tag side''),
then look for consistency with an invisible decay or a decay to a single photon of the other neutral $B$
(referred to as the ``signal side'').
The choice of reconstructing semileptonic \Bz decays on the tag side, with respect to fully-reconstructed \Bz final
states, is motivated by a higher reconstruction efficiency.  A disadvantage is the presence of the invisible neutrino, 
which prevents the exploitation of kinematic variables such as the reconstructed \Bz mass.
However, the background contamination is mitigated by the presence of a high momentum lepton.

We reconstruct $D^{*-}$ mesons in the final states 
$\Dzb\pi^{-}$ or $\Dm\piz$, with
\Dzb decays to $K^{+}\pi^{-}$, $K^{+}\pi^{-}\piz$, or $K^{+}\pi^{-}\pi^{+}\pi^{-}$,
and \Dm decays to $K^{+}\pi^{-}\pi^{-}$ or $\KS\pim$.
We identify $K^{+}$ candidates using 
Cherenkov-light information from the DIRC and energy-loss information (d$E$/d$x$) from the DCH and SVT.
The \KS mesons are reconstructed in the decay mode $\KS \to \pip\pim$, where the \pip\pim invariant mass lies in a
$\pm$25~\mevcc window around the nominal Particle Data Group (PDG) \KS mass~\cite{PDG}.
The \piz candidates are composed of pairs of photons observed in the EMC.  
Each photon must have a reconstructed energy above 30~\mev in the laboratory frame,
and the sum of their energies must be greater than 200~\mev.
The \piz candidates must have an invariant mass between 115 and 150~\mevcc.  A mass-constrained fit is
imposed on \piz candidates in order to improve the resolution on the reconstructed invariant mass of the parent $D$ meson.

Kaon and pion candidates are then combined to reconstruct $D^{(*)}$ mesons.
These are required to have an invariant mass within 60~\mevcc of their nominal 
PDG mass,
except for \Dzb decays with a \piz daughter, which must be within 100~\mevcc of the nominal \Dzb mass. 
Mass-constrained fits are applied to \Dzb and \Dm candidates in order to improve the measurement of the
momentum of each $D$.  
The difference in reconstructed mass between \Dstarm decay candidates and their $D$ daughters must be less than
175~\mevcc and greater than 137~\mevcc.
All $D^{(*)-}$ candidates must have a total
momentum between 0.5 and 2.5~\gevc in the CM frame.

Tracks selected as lepton candidates must pass either electron or muon selection criteria.
We identify electron candidates using energy and cluster shape information from the EMC, and Cherenkov angle information
from the DIRC.  Muon candidates are identified using information from the IFR and EMC.
Both electrons and muons must also have a momentum of at least 0.8~\gevc in the laboratory frame, and a minimum of 20 DCH measurements.

To further select $\Bz \to D^{(*)-}\ell^{+}\nu$ candidates, we require a $D^{(*)-}$ candidate and a lepton candidate 
to be consistent with production at a common point in space.
The decay vertex is reconstructed from a kinematic fit to all the candidate daughters, and a minimum
$\chi^{2}$ vertex probability of 0.001 is required.
We then calculate the cosine of the angle between the $D^{(*)-}\ell^{+}$ and the hypothesized \Bz candidate in the CM frame,
under the assumption that the only particle missing is a neutrino:
\begin{equation}
\hspace*{-0.2cm}\cos \theta_{B,D^{(*)-}\ell^{+}} = \frac{2\,E_{B} E_{{D^{(*)-}\ell^+}} -m^2_{{\B}} - m^2_{{D^{(*)-}\ell^+}}}
{2\,|\vec{p}_{B}| |\vec{p}_{{D^{(*)-}\ell^+}}|   }.
\end{equation}
The energy in the CM frame $E_{{D^{(*)-}\ell^+}}$ and mass $m_{{D^{(*)-}\ell^+}}$ of the $D^{(*)-}\ell^+$ combination are determined
from reconstructed momentum information, and $m_{B}$ is the nominal \Bz mass~\cite{PDG}.
The \Bz momentum $|\vec{p}_{B}|$ and energy $E_{B}$
in the CM frame are determined from beam parameters.
If our assumption that there is only one missing particle, a neutrino,
in the \Bz decays is incorrect,
$\cos \theta_{B,D^{(*)-}\ell^{+}}$ can fall outside the region $[-1,1]$.  
We require the $D^{(*)-}\ell^{+}$ combination to satisfy $-5.5 < \cos \theta_{B,D^{(*)-}\ell^{+}} < 1.5$.
The selected region allows for non-physical $\cos \theta_{B,D^{(*)-}\ell^{+}}$ values, accounting for detector energy and momentum resolution.
Moreover the asymmetric cut admits higher \Dstar mass states where additional decay products are lost. In the rest of
the analysis such products are not associated with the tag side decay chain but are considered as extra particles in the event.
When more than one $B^{0}\to D^{(*)-}\ell^{+}\nu$ candidate
is reconstructed in an event, the one with the highest vertex probability is taken.

\begin{figure}[!t]
\begin{center}
\includegraphics[width=8.5cm]{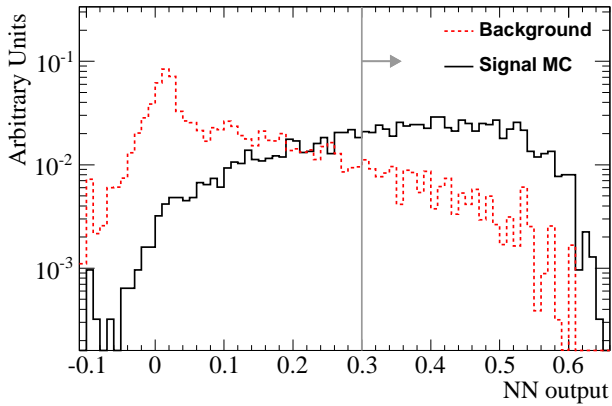}
\caption{
Distributions of the NN output for simulated \Bz $\to$ invisible events with a $D$ meson on the tag side.
The black solid line is the signal while the red dashed line is the background. The solid gray vertical line
defines the NN output signal region.
}
\label{fig:NNout}
\end{center}
\end{figure}

We consider events with no charged tracks besides those from the $\Bz \to D^{(*)-}\ell^{+}\nu$ candidate.
In order to reject background events where one charged or neutral particle is lost along the beam pipe,
the cosine of the polar angle of the missing momentum in the CM frame ($\cos\theta^{*}_{miss}$) is required to lie in the $[-0.9,0.9]$ range.
The missing 4-momentum due to unreconstructed particles is defined as the 
difference between the \FourS and the reconstructed tag side 4-momentum. In the \Btonunubg channel
the signal-side photon 4-momentum is also subtracted from the \FourS one.  

For the \Btonunub decay, in events where the $D$ meson on the tag side decays into $K^{-}\pi^{+}\pi^{-}$,
two additional selection criteria are applied. The first concerns the sum of the cosine of the angles between the kaon and two pions,
$\cos\theta_{K\pi_{1}}+\cos\theta_{K\pi_{2}}>-0.8$, while the second concerns the sum of the cosine of the angles between the lepton and the pions,
$\cos\theta_{\ell\pi_{1}}+\cos\theta_{\ell\pi_{2}}<0.8$. The main effect of this selection is the 
reduction of the background from misreconstructed $\epem \to \tau^{+}\tau^{-}$ events.

To reconstruct \Btonunubg events, one remaining photon candidate with energy 
greater than 1.2~\gev in the CM frame is also required. If the detected photon has an energy smaller than 
1.2~\gev in the CM frame, the event falls in the \Btonunub category and the neutral candidate is considered as an extra photon in 
the event. The choice of this cut generates a cross-feed between the two channels; MC simulation studies show that this has
a negligible effect on the final result. 

An artificial neural network (NN) is used to provide further discrimination between signal and background events. We use the TMVA software
package~\cite{tmva} and its multilayer perceptron implementation of a NN.
The architecture of the NN is composed of one input layer and one hidden layer.  These layers have
$V$ and $2V$ nodes, respectively, where $V$ is the number of the input variables.  
Samples that represent the signal and background components are given as input to the NN;
one half of each of these samples is used for the training while the other half is used as test. 
Once the NN has been trained, the output distributions for training 
and test samples are compared in order to check the presence of overtraining problems.
For the signal sample, MC simulation in which a generic semileptonic \B decay is generated and reconstructed is used.
Weighted off-peak data (composed of $\epem \to \ccbar, \uubar, \ssbar$, and $\tautau$ events, denoted as continuum background)
and MC simulated \BB events are used to describe the background contamination. 
Off-peak data are used 
to model continuum background, as the 
MC was found to incorrectly reproduce the cross-section of
two-photon fusion events, such as $\epem \to \epem\gamma\gamma\to\epem q\bar{q}/\tautau$. 
These events typically have decay products directed along the beam lines, and thus
outside the detector acceptance.

The variables used as input for the NN, common to the \Btonunub and \Btonunubg analyses, are:
1) $\cos \theta_{B,D^{(*)-}\ell^{+}}$; 2) the cosine of the angle in the CM frame between the thrust axis
(the axis along which the total longitudinal momentum of the event is maximized) and the $D^{(*)-}\ell^{+}$ pair momentum direction; 
and 3) the lepton momentum in the CM frame.
In the \Btonunub analysis, we additionally use 1$^{\prime}$) $M_{miss}^{tag}$ (defined as the invariant mass of the event after the 
$D^{(*)-}\ell^{+}$ pair is subtracted); 2$^{\prime}$) the $B$ meson vertex fit probability; 3$^{\prime}$) the ratio between the first and the zeroth-order $L$
momenta in the CM frame:
\begin{equation}
 L_{i}= \sum p\cos^{i}\theta,
\label{eq:legendre}
\end{equation}
where the sum is over extra tracks and neutrals and $\theta$ is computed with respect to the thrust axis;
4$^{\prime}$) the transverse momentum of the $D^{-}\ell^{+}$ pair in the CM frame; 5$^{\prime}$) the minimum invariant mass of any two charged tracks in the event;
and 6$^{\prime}$) the minimum invariant mass of any three charged tracks in the event.  Variables 4$^{\prime}$) -- 6$^{\prime}$) enter the NN only in the case of 
a reconstructed $\Bz \to D^{-}\ell^{+}\nu$ decay on the tag side.
In the \Btonunubg analysis, we additionally use 1$^{\prime\prime}$) the energy of the photon on the signal side evaluated in the laboratory frame and 
2$^{\prime\prime}$) $M_{miss}^{tag}$ (for $\Bz \to D^{-}\ell^{+}\nu$ reconstructed events only).

The selection on the output of the NN is optimized by minimizing the expected upper limit on the branching fraction, 
defined by a Bayesian approach as detailed later in this paper, under the hypothesis of observing zero signal events. This optimization is performed by using $B\bar{B}$ MC simulation and weighted off-peak data samples for the background estimation and the signal MC sample for the 
selection efficiency.
In Fig.~\ref{fig:NNout}, the output of the NN for simulated \Btonunub with a $D$ meson on the tag side
and the corresponding signal region are shown.

After the NN selection, the $D$ meson invariant mass ($m_{D}$) and the difference between the reconstructed $D^{*}$ invariant
mass and the PDG \Dz mass ($\Delta m$) are used to define a signal region (SR) and a side band region (SB) for 
the $D$ tag and \Dstar tag samples, respectively.
The SR is defined as a $\pm$15~\mevcc window around the PDG value for $m_{D}$ for the $\Bz \to D^{-}\ell^{+}\nu$ sample, and
as 0.139 $< \Delta m <$ 0.148~\gevcc for the $\Bz \to D^{*-}\ell^{+}\nu$ sample.  The excluded regions are used as the SB region. 

The total energy in the EMC computed in the CM frame and
not associated with neutral particles or charged tracks used in the $D^{(*)-}\ell^{+}$ reconstruction
is denoted as  $E_{\rm extra}$.  For \Btonunubg, the energy of the highest-energy photon remaining in the event
(the signal photon candidate) is also removed from the $E_{\rm extra}$ computation.
The $E_{\rm extra}$ signal region is defined by imposing an upper bound at 1.2~\gev.
In both \Btonunub and \Btonunubg samples, this variable is strongly peaked near zero for signal,
whereas for the background the distribution increases uniformly in the chosen signal region. 
Background events can, however, populate the low $E_{\rm extra}$ region, when charged or
neutral particles from the event are either outside the fiducial volume of the detector or are unreconstructed 
due to detector inefficiencies.  Contributions from misreconstructed \piz decays usually populate the high $E_{\rm extra}$ region.

Using detailed Monte Carlo simulations of \Btonunub and \Btonunubg events,
we determine our signal efficiency to be $(17.8 \pm 0.2) \times 10^{-4}$
for \Btonunub and $(16.0 \pm 0.2) \times 10^{-4}$ for \Btonunubg, where the uncertainties 
are statistical. These efficiencies are enhanced by a factor 8.5\% and 11\%, respectively, with
respect to the previous analysis \cite{prevprl}. The background
selection efficiencies (evaluated in $B\bar{B}$ MC plus off-peak data)
are $4.16 \times 10^{-8}$ and $1.32 \times 10^{-9}$ for the invisible
and invisible$+\gamma$ decay, respectively.  These can be compared with the
background selection efficiencies in the previous analysis, which were
$2.79 \times 10^{-7}$ and $4.96\times10^{-8}$, respectively.

We construct probability density functions (PDFs) for the $E_{\rm extra}$ distribution 
for signal ($\mathcal{P}_{\rm sig}$) and background ($\mathcal{P}_{\rm bkg}$) using 
detailed MC simulation for signal and data from the $m_{D}$ and $\Delta m$ sidebands for background.
The two PDFs are combined into an extended maximum likelihood function 
$\mathcal{L}$, defined as a function of the free parameters $N_{\rm sig}$ and $N_{\rm bkg}$,
the number of signal and background events, respectively:
\begin{IEEEeqnarray}{rCl}
 \mathcal{L}(N_{\rm sig},N_{\rm bkg}) & = & \frac{\left[(1-z_{\rm sig})N_{\rm sig}+(1-z_{\rm bkg})N_{\rm bkg}\right]^{N_{1}}}{N_{1}!} \nonumber\\
&\times & e^{-\left[(1-z_{\rm sig})N_{\rm sig}+(1-z_{\rm bkg})N_{\rm bkg}\right]}\nonumber\\
&\times & \prod_{i=1}^{N_{1}}\left[ \mathcal{P}_{\rm sig}(E_{{\rm extra},i}|\vec{p}_{\rm sig})\frac{(1-z_{\rm sig})N_{\rm sig}}{N_{1}}\right. \nonumber \\
&+&\left.  \mathcal{P}_{\rm bkg}(E_{{\rm extra},i}|\vec{p}_{\rm bkg})\frac{(1-z_{\rm bkg})N_{\rm bkg}}{N_{1}} \right]\nonumber\\
&\times & \frac{\left(z_{\rm sig}N_{\rm sig}+z_{\rm bkg}N_{\rm bkg}\right)^{N_{0}}}{N_{0}!}\nonumber \\
&\times&     e^{-\left(z_{\rm sig}N_{\rm sig}+z_{\rm bkg}N_{\rm bkg}\right)}.
\label{eqn:lik}
\end{IEEEeqnarray}
The photon reconstruction has a detection lower energy
bound of 30~\mev, and as a consequence, the $E_{\rm extra}$ distribution is not continuous.
To account for this effect, the likelihood in Eq.~\ref{eqn:lik} is composed of two distinct parts, one for $E_{\rm extra}>30$~\mev
and one for $E_{\rm extra}=0$~\mev. In 
the likelihood function,
$z_{\rm sig}$ and $z_{\rm bkg}$ are the fractions of events with
$E_{\rm extra}=0$~\mev for signal and background, respectively, and $\vec{p}_{\rm sig}$ and $\vec{p}_{\rm bkg}$ are the vectors of
parameters describing the signal and background PDFs, a kernel-based PDF~\cite{KEYS} and a second-order polynomial, respectively.
The fixed parameters $N_{0}$, $N_{1}$, and $E_{{\rm extra},i}$ are, respectively, the number of events with $E_{\rm extra}=0$~\mev,
the number of events with $E_{\rm extra}>30$~\mev, and 
the value of $E_{\rm extra}$ for the $i^{th}$ event.  

\begin{figure}[!t]
\begin{center}
\begin{minipage}[h]{4.25cm}
\includegraphics[width=4.25cm]{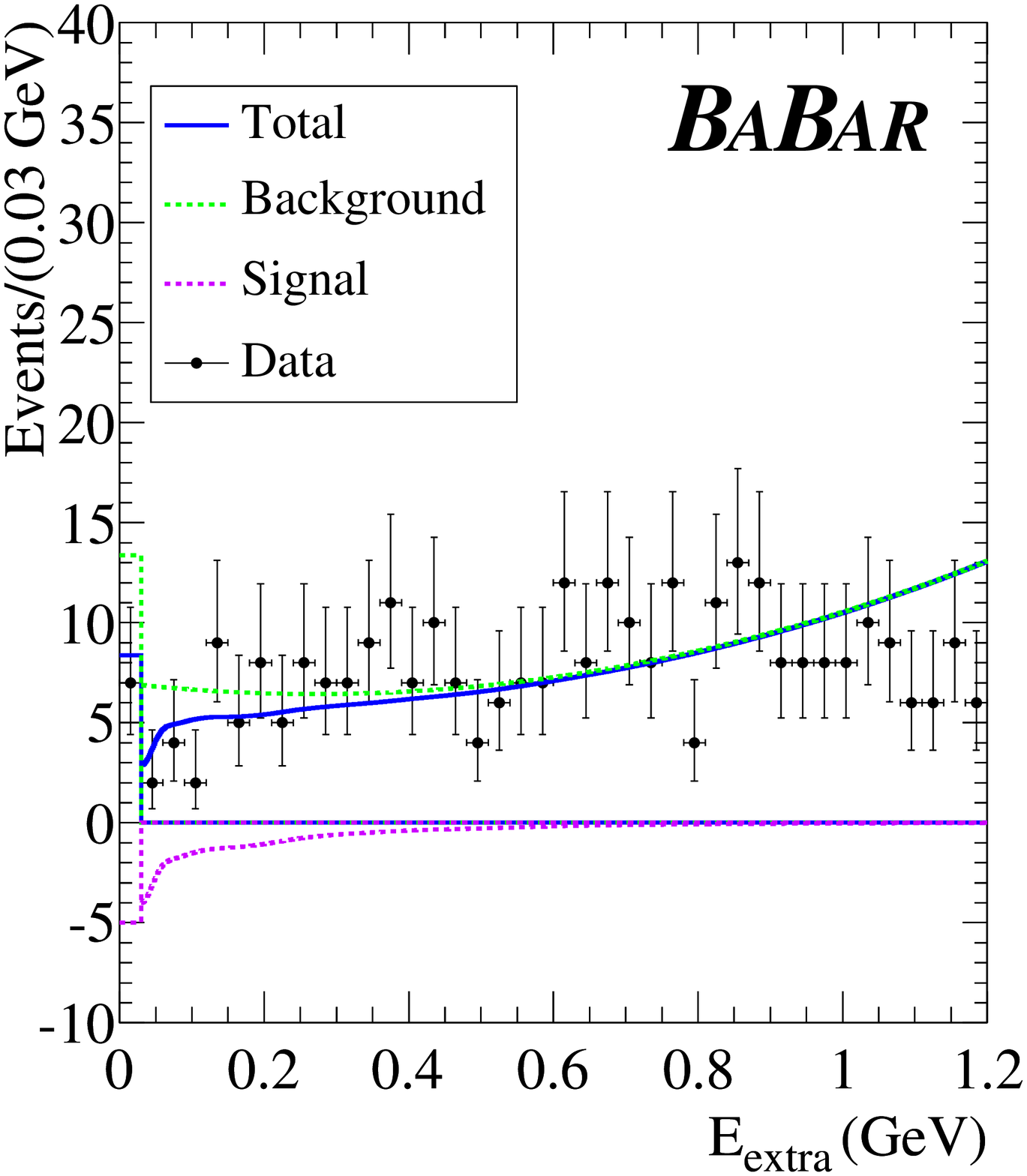}
\end{minipage}
\begin{minipage}[h]{4.25cm}
\includegraphics[width=4.25cm]{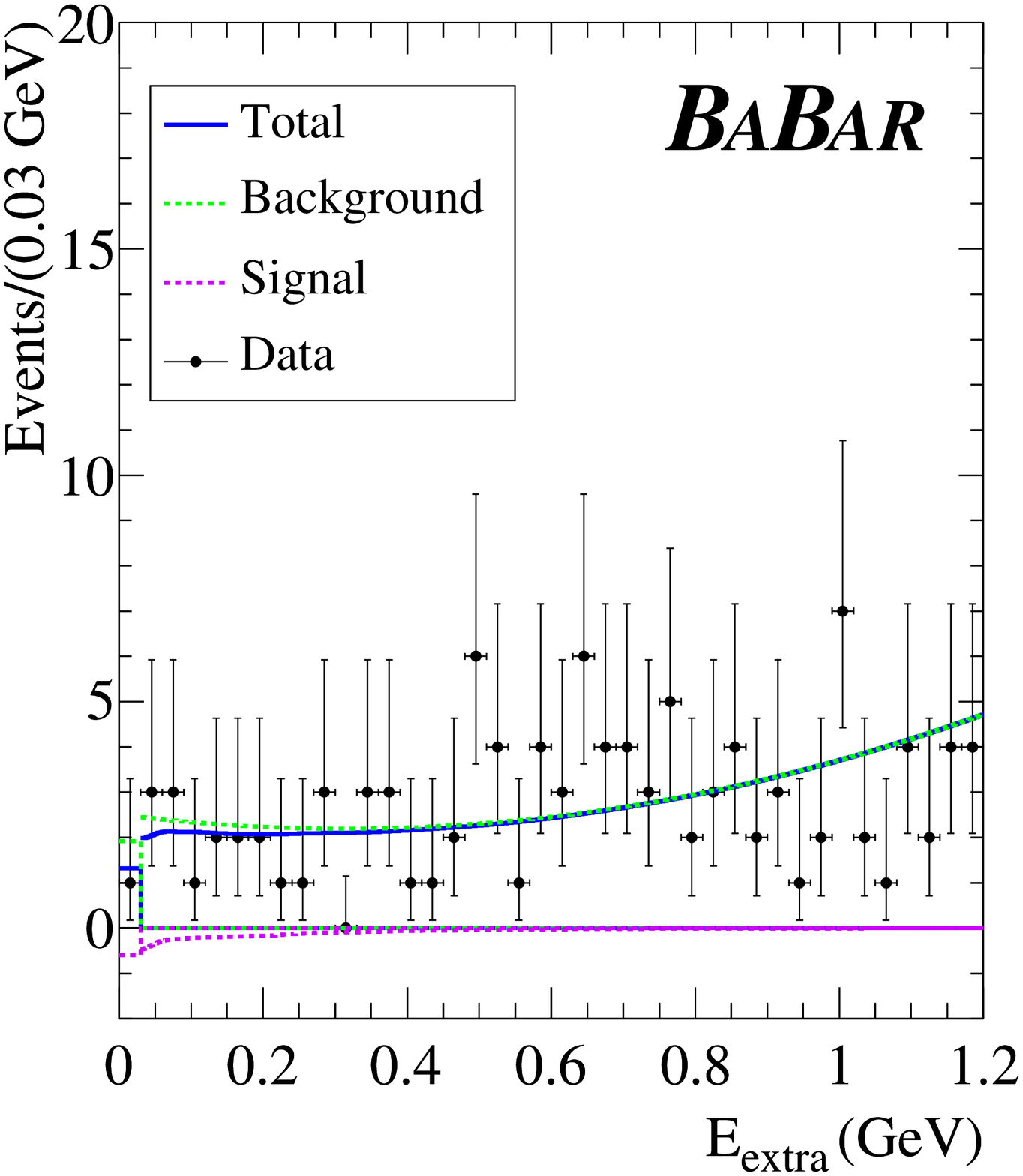}
\end{minipage}
\caption{
Results of the maximum likelihood fit of $E_{\rm extra}$ for \Btonunub (left) and \Btonunubg (right).
}
\label{fig:etotleft}
\end{center}
\end{figure}

The negative log-likelihood
is then minimized with respect to 
$N_{\rm sig}$ and $N_{\rm bkg}$ in the data sample.  
The resulting fitted values for $N_{\rm sig}$ and $N_{\rm bkg}$ are given in Table~\ref{tab:yields}.  
Figure~\ref{fig:etotleft} shows the $E_{\rm extra}$ distributions for 
\Btonunub and \Btonunubg with the fit superimposed.

\begin{table}[hb]
  \caption{Fitted yields of signal and background events in data.  The uncertainties are statistical.}
  \label{tab:yields}
  \begin{center}
    \begin{tabular}{lcc}
      \hline
      \hline
      Mode       & $N_{\rm sig}$ & $N_{\rm bkg}$ \\
      \hline
      \rule{0pt}{12pt}\Btonunub  & $-22 \pm 9$    & $334 \pm 21$  \\
      \rule{0pt}{12pt}\Btonunubg & $-3.1 \pm 5.2$ & $113 \pm 12$  \\
      \hline
      \hline
    \end{tabular}
  \end{center}
\end{table}

The fitted signal yields are used to determine the decay branching fractions ($\mathcal{B}$), 
which are defined as:
\begin{equation}
 \mathcal{B} \equiv \frac{N_{\rm sig}}{\varepsilon \times N_{B\bar{B}}} \, ,
\label{eq:BR}
\end{equation}
where $\varepsilon$ is the total signal efficiency, corrected for data-MC discrepancies (described below), and
$N_{B\bar{B}}$ is the number of produced \BB pairs.

The systematic uncertainty on the signal efficiency is dominated
by data-MC discrepancies in the distribution of the variables used as input to the NN. 
This results in relative uncertainties of 6.1\% and 8.2\% for \Btonunub and \Btonunubg, respectively.
This uncertainty is evaluated using the hypothesis that the data-MC agreement could reduce the discriminating
power of each input variable. In order to make the signal distributions more background-like,
in the signal sample we apply a Gaussian smearing to each of the NN input variables, where the smearing parameters are 
evaluated by comparing the difference in the root-mean-square of the signal and background shapes.
With this method correlations between variables are not considered but specific studies have indicated
that the impact of the correlations is negligible.
The NN output selection is then applied to this new sample and the difference between the nominal signal
efficiency and this new efficiency is used as the systematic uncertainty.

Another important contribution is due to the estimation of the efficiency on the tag side reconstruction
(3.5\% for both channels). For this purpose,
data and MC samples in which a \Bz and a \Bzb are both reconstructed as decays to $D^{(*)}\ell\nu$ in the same event (``double tag'' events) are used.
The square root of the ratio between the number of the selected double tag events in data and in MC simulation is 
0.928 (0.824) for events with $\Bz\to D^{(*)}\ell\nu$ on the tag side; these ratios are used to correct the efficiency.  The 
propagation of the 
statistical errors on the correction factors is used
as a systematic uncertainty on the signal efficiency. 

Other contributions to the systematic uncertainty on the signal efficiency come from the choice of the preselection
criteria and from the SR definition of $m_{D}(\Delta m)$. The first effect is evaluated by applying a Gaussian smearing to the 
variables involved
($\cos\theta^{*}_{miss}$, $\cos\theta_{K\pi_{1}}+\cos\theta_{K\pi_{2}}$ and $\cos\theta_{\ell\pi_{1}}+\cos\theta_{\ell\pi_{2}}$). 
The variation on the signal efficiency is then used as a systematic uncertainty.  As was done for the NN, this uncertainty is evaluated using the hypothesis that the discrimination
power of each variable is reduced.  The second effect is evaluated by changing each of the bounds of the SR definition by a 
value $\delta$
(3~\mev for $m_{D}$ and 1.5~\mev for $\Delta m$), which is half of the $m_{D}$/$\Delta m$ resolution as evaluated in data.
The relative maximum variation in efficiency is then used as a systematic uncertainty.

An additional source of systematic uncertainty is determined for the \Btonunubg decay in order to account for detector inefficiency in the single photon reconstruction.  
This is evaluated by comparing the data and MC \piz reconstruction efficiency 
in $\tau\to\rho(\pi^{\pm}\piz)\nu$ decays, where the total number of produced \piz in the selected sample is determined from
the branching fraction of the specific $\tau$ decay~\cite{PDG}. Then the ratio between the two efficiencies,
combined with the error on the $\tau$ decay branching ratio, is used to extract a systematic error
for the single photon reconstruction efficiency.

The total systematic uncertainty on the signal selection efficiency is 7.7\% for \Btonunub decay and 9.5\% for \Btonunubg decay.

The systematic uncertainty on the number of signal events is dominated by the parametrization
\begin{table}[ht]
  \caption{Summary of the systematic uncertainties.}
  \label{tab:totsys}
  \begin{center}
    \begin{tabular}{ccc}
      \hline
      \hline
      \rule{0pt}{12pt} Source & \Btonunub & \Btonunubg\\
      \hline
      \rule{0pt}{1pt}\\
      \multicolumn{3}{c}{Normalization Errors}\\
      \hline
      $B$-counting & $0.6\%$ & $0.6\%$\\
      \hline
      \rule{0pt}{1pt}\\
      \multicolumn{3}{c}{Efficiency Errors}\\
      \hline
      Tagging Efficiency & $3.5\%$ & $3.5\%$\\
      $m_{D}$ $(\Delta m)$ Selection & $1\%$ & $1.3\%$\\
      Preselection   & $3\%$ & $2.4\%$\\
      Neural Network & $6.1\%$ & $8.2\%$\\
      Single Photon & -- & $1.8\%$\\
      \hline
      TOTAL & $7.7\%$ & $9.5\%$\\
      \hline
      \rule{0pt}{1pt}\\
      \multicolumn{3}{c}{Yield Errors (events)}\\
      \hline
      Background Param. & $15.8$ & $6.5$\\
      Signal Param. & $2.0$ & $1.2$\\
      Fit Technique  & -- & $1.0$\\
      $E_{\rm extra}$ Shape & $0.1$ & $1.8$\\
      \hline
      TOTAL & $15.9$ & $6.9$\\
      \hline
      \hline
    \end{tabular}
  \end{center}
\end{table}
of the background $E_{\rm extra}$ distribution. A maximum likelihood fit of $E_{\rm extra}$ with the background parameters varied according to their
statistical error and correlations is performed. For each parameter the difference in the fitted signal yield with respect to the nominal value is used as a 
systematic uncertainty. 
Other contributions to the signal yield systematic uncertainty
come from the signal shape parametrization and from the use of the data SB for the determination of the background shape.
The first is evaluated as the difference between the fitted yield with the polynomial shape and an alternative exponential shape.
The latter, computed as the difference in the $E_{\rm extra}$ shape between the SR and SB, is parametrized with a first-order 
polynomial
using the charge-conservation violating \Bptonunub$\!\!(\!+\gamma)$ control sample discussed below.
This parametrization is used to weight the background shape, and the difference in the fitted
yield is used as a systematic uncertainty. Another contribution for the \Btonunubg decay is due to a small bias observed in MC 
studies of the yield extraction.
The total systematic uncertainties on the signal yield are 16 and 7 events for \Btonunub and \Btonunubg, respectively.

For the systematic contribution due to the uncertainty on the estimation of the total number of \BB events in the data sample, the procedure adopted is described in
Ref.~\cite{bcount}
and the resulting uncertainty is $0.6\%$.  The systematic uncertainties are summarized in Table~\ref{tab:totsys}.

A Bayesian approach is used to set 90\% confidence level (CL) upper limits on the branching fractions for \Btonunub and \Btonunubg.
Flat prior probabilities are assumed for positive values of both branching fractions.  Gaussian likelihoods are adopted for signal yields.
The Gaussian widths are fixed to the sum in quadrature of the statistical and systematic yield errors. We extract a posterior PDF using Bayes' theorem,
including in the calculation the effect of systematic uncertainties associated with the efficiencies and the normalizations, modeled by Gaussian PDFs.
Given the observed yields in Table~\ref{tab:yields}, the 90\% confidence level upper limits are calculated, after the marginalization of the posterior PDF, by
\begin{equation}
\int_{0}^{UL} \mathcal{P}(\mathcal{B})d\mathcal{B}{\bigg/}\int_{0}^{\infty} \mathcal{P}(\mathcal{B})d\mathcal{B}=0.9.
\end{equation}
The resulting upper limits on the branching fractions are
\begin{eqnarray}
\mathcal{B}(\Btonunub) & < & 2.4 \times 10^{-5}\nonumber\\
\mathcal{B}(\Btonunubg) & < & 1.7 \times 10^{-5}\nonumber
\end{eqnarray}
at 90\% CL. 
In order to cross-check the results of the analysis, we also search for the charge-conservation violating modes \Bptonunub and \Bptonunubg.  
We check that their resulting signal is consistent with zero.  For these modes, we reconstruct 
\Bpm $\to D^0\ell\nu X^0$, where $X^0$ can be a photon, \piz, or nothing.  The \Dz is reconstructed in the same three decay modes
as in $\Bz \to D^{(*)-}\ell^{+}\nu$, and similar criteria are enforced for the reconstructed charged $B$ as for the neutral $B$ modes.  
The resulting fitted values of $N_{\rm sig}$ 
are $-4.3\pm3.8$ (stat.)~for
\Bptonunub
and $-7.9\pm8.3$ (stat.)~for
\Bptonunubg, which are both consistent with zero within 1.1 standard deviations.  

In summary, we obtain improved limits on branching fractions for \Bz decays to an invisible final state and for \Bz decays to 
invisible$+\gamma$.
The upper limits at 90\% confidence level are
$2.4 \times 10^{-5}$ and $1.7 \times 10^{-5}$ for the \Btonunub and \Btonunubg
branching fractions, respectively.
The latter limit assumes a photon momentum distribution predicted by the constituent 
quark model for $\Bz \to \nu\bar{\nu}\gamma$ decay~\cite{nunugam},
whereas the \Bz $\to$ invisible limit is not decay-model dependent.
These limits supersede our earlier results~\cite{prevprl}, which used a small fraction of our present dataset.

We are grateful for the excellent luminosity and machine conditions
provided by our \pep2\ colleagues, 
and for the substantial dedicated effort from
the computing organizations that support \babar.
The collaborating institutions wish to thank 
SLAC for its support and kind hospitality. 
This work is supported by
DOE
and NSF (USA),
NSERC (Canada),
CEA and
CNRS-IN2P3
(France),
BMBF and DFG
(Germany),
INFN (Italy),
FOM (The Netherlands),
NFR (Norway),
MES (Russia),
MICIIN (Spain),
STFC (United Kingdom). 
Individuals have received support from the
Marie Curie EIF (European Union)
and the A.~P.~Sloan Foundation (USA).


\begin{thebibliography}{99}
\bibitem{conjugates}
Charge-conjugate decay modes are implied throughout this paper.
\end{thebibliography}
\end{document}